\documentclass[12pt,preprint]{aastex}

\def\leaderfill{\leaders\hbox to 1em{\hss . \hss}\hfill}

\def\etal         {{et al.~}}

\usepackage{epsfig}
\usepackage{epsf}

\begin{document}
 
\def\today{\number\year\space \ifcase\month\or  January\or February\or
        March\or April\or May\or June\or July\or August\or
September\or
        October\or November\or December\fi\space \number\day}
\def\fraction#1/#2{\leavevmode\kern.1em
 \raise.5ex\hbox{\the\scriptfont0 #1}\kern-.1em
 /\kern-.15em\lower.25ex\hbox{\the\scriptfont0 #2}}
\def\spose#1{\hbox to 0pt{#1\hss}}
\def\simlt{\mathrel{\spose{\lower 3pt\hbox{$\mathchar''218$}}
     \raise 2.0pt\hbox{$\mathchar''13C$}}}
\def\simgt{\mathrel{\spose{\lower 3pt\hbox{$\mathchar''218$}}
     \raise 2.0pt\hbox{$\mathchar''13E$}}}

\title{Formation of Carbon Dwarfs}
\author{Charles L. Steinhardt \\ Dimitar D. Sasselov}
\affil{Department of Astronomy, Harvard University, 60 Garden St., Cambridge, 
MA 02138}

\begin{abstract}
Since low-mass main sequence stars do not produce carbon, the large number of carbon dwarfs recently discovered using the Sloan Digital Sky Survey presents a puzzle requiring a theoretical explanation.  We consider the formation of dwarf carbon stars via accretion from a carbon asymptotic giant branch (AGB) companion in light of the 107 object sample of Downes et al. (2004).  This sample is now large enough to distinguish between carbon giants and dwarfs via comparison of a composite spectrum to theoretical atmospheric models.  Carbon dwarfs of spectral type M are indeed main sequence M dwarfs with enhanced metallicity and carbon abundance.  We also calculate the predicted abundance of both M and of F/G carbon dwarfs, and show that the latter should be falsifiable in the near future.  
\end{abstract}





\maketitle


\section{Introduction}
\label{sec:intro}

A carbon star is a star with carbon to oxygen ratio (C/O) greater than one and has a distinct atmospheric chemistry that yields a spectrum easily distinguishable from a typical (C/O $<1$) star.  Observationally, it has proven much easier to detect large and intrinsically luminous asymptotic giant branch (AGB) stars with C/O$>1$ than it is to detect the much fainter carbon dwarfs at the low-mass end of the main sequence.  As a result, it was generally assumed that carbon dwarfs were exotic objects, and that most carbon stars were on the asymptotic giant branch.  Only recently has it become apparent that main sequence carbon dwarfs comprise not only a sizable population, but perhaps even the majority of all carbon stars \cite{Green1992,DeKool1995,Green2000,Margon2004}.  Since low-mass main sequence stars do not produce carbon in their cores and there is less carbon than oxygen in the interstellar medium, it is natural to ask how these carbon dwarfs form.

The first main sequence carbon star detected, G77-61 \cite{Dahn1977}, is a radial velocity binary with a cool white dwarf companion \cite{Dearborn1986}, and the carbon dwarf stars PG 0824+289 \cite{Heber1993} and CBS 311 \cite{Liebert1994} are spectroscopic binaries with hot white dwarf companions.  These authors concluded that carbon dwarfs may have formed via accretion from a carbon AGB companion.  De Kool and Green (1995)\nocite{DeKool1995} showed that such a model indeed produces a carbon dwarf population more numerous than that of carbon AGB stars.  

We consider the remaining issues with this model in light of new observations by the Sloan Digital Sky Survey.  What remains to be done in order to establish that dwarf carbon stars are binary companions of former AGB stars?  First, while carbon dwarfs have spectral types consistent with the late main sequence, it has not yet been possible to precisely determine their masses.  If, for example, a solar mass carbon dwarf appeared as a spectral type M due to unusual, poorly-resolved spectral lines, then simulations would have assumed the star to be fully convective (i.e., the entire star has C/O$>1$) while for a solar mass star C/O need only be $>1$ in a thin stellar envelope.  Second, while a few carbon dwarfs with white dwarf companions are known, it has not been possible to find a companion for every known carbon dwarf.  Combined with the small sample size of observed binary systems, this is not yet compelling evidence for the production of carbon stars by binary mass transfer; after all, 42\% of stellar systems with an M star as their primary are binary \cite{Fischer1992}.  Certainly additional observations showing that carbon dwarfs indeed have white dwarf companions would be compelling evidence for the binary mass transfer model.  We can also search for other predictions that might provide for stringent tests of this model.

In {\S~\ref{sec:spectra}} we analyze spectra produced by the Sloan Digital Sky Survey of carbon dwarf candidates and attempt to quantitatively determine the properties of this population.  In {\S~\ref{sec:dynamics}} we consider the dynamics of a binary system involving an AGB and an M star and whether the endstate of such a system might be a dwarf carbon star, as well as whether such carbon dwarfs would be produced in the observed frequency.  In {\S~\ref{sec:GandS}}, we consider the new predictions that such a model makes with an eye towards verifiability.  Finally, in {\S~\ref{sec:conclude}}, we consider the additional science that would be made possible by the confirmation of this theory.

\section{Analysis of SDSS Spectra}
\label{sec:spectra}
Recent observations from the Sloan Digital Sky Survey (SDSS) have been able to observe stars as dim as 23rd magnitude \cite{EDR,DR1,DR2} and therefore include a large sample of main sequence dwarfs.  Because a carbon dwarf spectrum does not exhibit strong TiO lines, carbon dwarfs are uniquely identifiable using the five filters applied by SDSS.  In addition to the theoretically well understood carbon AGB stars comprising approximately 10\% of the total AGB population, SDSS also finds that a surprisingly large 0.3-0.4\% of main sequence M stars show spectral features indicative of C/O$>1$ \cite{Schroeder2003}.  In addition, the survey has taken low-resolution (69 km/s pixel) spectra of dozens of carbon dwarf candidates, including one sample of 103 dwarf carbon stars \cite{Margon2002,Margon2004}.  Schroeder (2003)\nocite{Schroeder2003} showed that the sample is indeed distinct from low-gravity giant carbon stars by comparing composite SDSS spectra of giants and dwarfs.  We determine the parameters of these carbon dwarfs by matching their spectra to synthetic, simulated ones.

Our model atmospheres for these synthetic spectra are based on using an updated version of the ATLAS code \cite{Kurucz1992} and are built on the four standard assumptions: hydrostatic equilibrium, plane-parallel geometry, radiative equilibrium and local thermodynamic equilibrium.  The original ATLAS code has been modified to simulate a carbon-dominated atmosphere and uses a full set of carbon chemistry line lists as described by Allard et al. (2000)\nocite{Allard2000}.  We calculate the emergent synthetic spectrum for a range of atmospheric parameters (T$_{eff}$, $\log g$) in two main domain: those of a G-dwarf (solar analog) and of an M-dwarf. For each domain the carbon abundance is increased until the C$_2$ and CN bands appear in the optical region and match in strength the observed ones.

\begin{figure*}
\begin{center}
\includegraphics[width={0.9\columnwidth}]{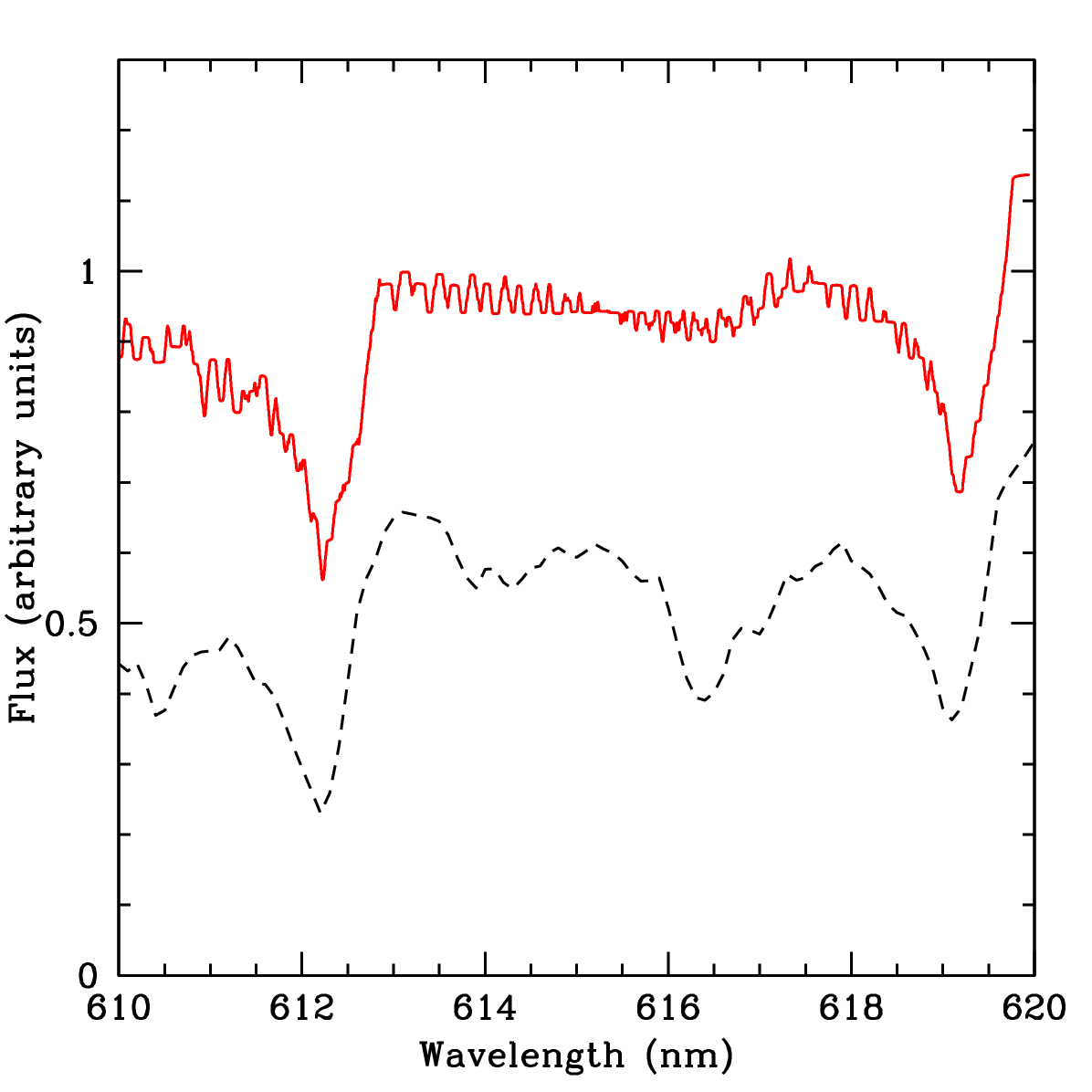}%
\end{center}
\caption{A comparison of the simulated M dwarf spectrum with enhanced C/O ratio (top, solid red line) to the composite observed carbon M dwarf spectrum (bottom, dashed black line), with the relative flux shown in arbitrary units.  Both have similar spectral line ratios, including the two prominent C$_2$ lines.  The $\lambda$6191 band is a luminosity discriminant between dwarfs and giants \cite{Green1992b}.  All strong spectral features are reproduced by the models; as is common with very cool stars, a number of weaker, unidentified spectral lines are often exhibited.}
\label{fig:modelcompare}
\end{figure*}

For both G- and M-dwarfs, as expected, the optical spectra attain the features of a carbon star spectrum when the carbon to oxygen ratio exceeds 1.05.  Both spectra were compared to the composite carbon M dwarf spectrum produced by Schroeder (2003)\nocite{Schroeder2003}.  Poor signal-to-noise in individual stellar spectra required the use of a composite spectrum, but the apparent homogeneity of the sample reported by Schroeder (2003)\nocite{Schroeder2003} implies that a comparison between the stacked spectrum and synthetic model spectra is meaningful.  The composite spectrum is a much better match with the spectrum of the main sequence M dwarf with enhanced C/O ratio than with the simulated carbon G dwarf.  Knapp (private communication) reports both carbon G dwarfs and carbon M dwarfs in SDSS, but the composite spectrum is dominated by M dwarfs.  In Figure \ref{fig:modelcompare}, we show that the composite spectrum is a match for the simulated main sequence M dwarf with enhanced C/O ratio.

The spectra of the observed objects were also inspected for atomic absorption lines: particularly prominent are the Na~{\small I} and K~{\small I} resonance doublets, as well as Ca~{\small I} and Mg~{\small I} lines, and the Ca~{\small II} triplet.  Matching these, as well as the Balmer H$\alpha$ and H$\beta$ lines, with the synthetic spectra allows a rough temperature separation (G vs. M dwarf) and luminosity separation (from pressure broadening).  A more difficult question is the overall metal content of these stars; the low resolution and S/N of the spectra prevent a detailed analysis; a comparison of our spectra with the analysis of two extremely metal poor carbon dwarfs by Cohen et al. (2004)\nocite{Cohen2004} indicates that our objects are not metal poor as a sample.  This is also true for the earlier sample provided by Norris et al. (2001)\nocite{Norris2001}.

We can conclude that the simulated carbon M dwarf spectra are a very good match for the composite spectra.  Therefore, the population flagged by SDSS is indeed composed of M stars with C/O$>1.05$ in the stellar photosphere.  

\section{Carbon Dwarfs as Post Mass-Transfer Binaries}
\label{sec:dynamics}

Analysis of SDSS spectra of dwarf carbon stars as done in {\S~\ref{sec:spectra}} shows that indeed at least 0.3-0.4\% of M stars on the main sequence are carbon stars and that those stars show significant enhancement in their carbon content.  Our goal is to explain how carbon dwarfs can comprise such a sizeable fraction of low mass main sequence stars.  When carbon dwarfs were considered exotic objects, a likely candidate for their formation was accretion from material originating in the photosphere of carbon-enriched AGB stars.  It is therefore natural to consider whether a more sizeable population might still be formed by mass transfer from AGBs.

The sole known class of stars for which C/O$>$1 is common in their photospheres are AGB stars.  They have a carbon-oxygen core surrounded by a hydrogen-burning layer and a helium-burning layer, and a convective zone that reaches into the layer of helium burning.  However, during a thermal pulse, material from the carbon layer above the core will be dredged up and enter the convective zone, and as a result the photosphere will be carbon enriched.  Since the material is composed of 22\% carbon and only 2\% oxygen by mass \cite{Groenewegen1993}, the photosphere can eventually attain C/O$>$1.  As a result, between 5 and 15 percent of AGB stars are carbon stars.
 
In addition, an AGB loses mass rapidly, in some cases as quickly as $10^{-4}M_\odot/\textrm{yr}$.  Over the AGB lifetime of $10^5\textrm{ yr}$, as much as 1-2$M_\odot$ may be ejected from the AGB.  As a result, de Kool and Green (1995)\nocite{DeKool1995} have proposed that dwarf carbon stars might be the end product of a binary system involving an AGB carbon star and an M dwarf.  Now that we have established that at least a factor of ten enhancement of carbon in the M star is required and that according to SDSS approximately 0.3-0.4\% of M stars show such enhancement, we can consider whether these numbers are consistent with such an explanation.

In order for this model to produce carbon M dwarfs of the type and in the quantity observed, we must show two things.  We must show that at the conclusion of the AGB phase of the primary star, the M dwarf will be left with a C/O ratio that matches the observations in the previous section.  We must also show that the rate of such binary systems yields the eventual 0.3-0.4\% fractional abundance of carbon stars among main sequence M dwarfs.

As the primary star becomes an AGB, it will overflow its Roche lobe and the binary system will enter a common envelope phase \cite{Iben19xx}.  During the common envelope phase, there is a net outflow from the AGB of as much as $10^{-4}M_\odot/\textrm{yr}$ with a typical velocity on the order of $20\textrm{ km/s}$.  We take the Bondi-Hoyle approximation that all gas passing within a cylindrical radius
\begin{equation}
r_{acc} = \frac{2GM}{v^2}
\end{equation}
will be accreted by the secondary star.  In the case of an M dwarf of mass $0.30 M_\odot$, this yields $r_{acc} = 1.3 \textrm{ AU}$.  For a binary system with separation $1 \textrm{ AU}$, $r_{acc}$ would subtend an angle of 53 degrees, so the secondary will accrete material within a solid angle of 2.5 steradians.  Thus, in the Bondi-Hoyle approximation, the mass $M_d$ of the dwarf changes as
\begin{equation}
\dot{M}_{d} \approx -0.20\dot{M}_{\textrm{AGB}}.
\end{equation}
If the orbit is not circular, as has been demonstrated for many AGB models \cite{Jorissen1998,Jorissen2009}, the actual mass transfer will be different but of a similar magnitude.

As discussed in {\S~\ref{sec:spectra}}, carbon M dwarfs show evidence of a range of metallicities, but as a whole are not metal poor.  This is not surprising, given both that the elemental abundances in the photosphere of a typical M dwarf are solar \cite{Ludwig2002} and that their low intrinsic luminosity restricts the SDSS sample to observations of nearby dwarfs.  So, we can take the initial abundances in the secondary star photosphere as \cite{Grevesse2004} $\textrm{log C}/\textrm{log H} = -3.61 \pm 0.05, \textrm{log O}/\textrm{log H} = -3.34 \pm 0.05$.  

Since it is typical for both stars in a binary system to have formed simultaneously, if we take solar composition for the M dwarf, we should take the same initial composition for the AGB progenitor.  Sivarani et al. (2004)\nocite{Sivarani2003} observed the metal-poor carbon AGB star CS 29497-030 to have $[Fe/H] = -2.8, [C/Fe] = +2.38,$ and $[O/Fe] = +1.67$.  Ohnaka et al. (2000)\nocite{Ohnaka2000} (see also \cite{Lambert1984}) find the abundances in two carbon stars TX Psc and V Aql with $[Fe/H] \sim -0.5$ and $-0.7$ as $[C/H] = 0.16$ and $0.13$ and $[O/H] = -0.14$ and $-0.36$, respectively.  If the surface gravity is lower than solar in TX Psc, Ohnaka et al. calculate an additional slight enhancement in C/O.  Note that all of these differences in observed abundances between individual carbon stars are insignificant compared to the effect discussed in this paper.  All of the above abundances are given in dex relative to Anders and Grevesse (1989)\nocite{Anders1989}.

As expected there is a metallicity dependence in $[C/H]$ and $[O/H]$, but for a star with $[Fe/H] = 0$, we can estimate from the above that $[C/H] \sim 0.75$ and $[O/H] \sim 0.35$.  Since the observed carbon AGB stars may be at different stages and both the C/O ratio and the mass loss rate increase with time, this should be a lower bound on both the C/H and C/O ratios.  These ratios in material accreted by the M dwarf companion are likely considerably greater than these bounds.

If the AGB indeed loses a total of $1 M_\odot$ over its lifetime, then using the initial abundances for a main sequence dwarf with mass $0.30 M_\odot$, the final mass of the secondary will be $M \approx 0.50 M_\odot$.  Its abundances will then be
\begin{equation}
[C/H]_d \sim 0.45,
\end{equation}
\begin{equation}
[O/H]_d \sim 0.17.
\end{equation}
This results in C/O $> 1.02$ as our lower bound.  We have not applied the surface gravity correction (which, if analogous to TX Psc, would result in C/O $> 1.05$), but the uncertainties in $[C/H]$, $[O/H]$, and $[Fe/H]$ for carbon AGB stars are much larger than this correction.  

As shown in {\S~\ref{sec:spectra}}, the C/O ratio needs to be enhanced until C/O $>1$, or around a factor of five, to produce a carbon star.  So we see that this model there is enough material accreted from the AGB and with the proper abundance to produce a carbon star of the type imaged by SDSS.  
A star with C/O very close to 1 will be an S star, showing prominent ZrO lines.  Because the atmospheric chemistry is so sensitive to the C/O ratio around 1, C/O = 1.02 is large enough to produce a carbon star rather than an S star.  However, it should be noted that if further investigation into the elemental abundances of carbon AGB stars reduce our lower bound, this might present a problem for the mass transfer model of carbon M dwarf formation.

Now, consider the rate at which M stars are the secondary in a binary system in which the primary will undergo an AGB phase.  In this case, we are interested in systems composed of a secondary of mass $M_s \sim 0.1-0.4 M_\odot$ and a primary of mass $1.5 < M_p/M_\odot$.  So, such a system has $q \equiv M_s/M_p < 0.4$, and a typical $q$ will be between $0.05$ and $0.2$.  There are considerable difficulties in determining the frequency of binary systems with different mass ratios, particularly for small values of $q$ because the corrections for detection biases can be larger than the uncorrected signal(cf. Latham 2002\nocite{Latham2002}).  Indeed, surveys have found that there is a peak, with the frequency decreasing with smaller $q$ \cite{Duquennoy1991}, that there are two peaks \cite{Latham2002}, or that the distribution is close to flat \cite{Mazeh1992}.  It is also possible that some of the differences between these samples are a result of examining different ranges for the mass of the primary.  

For each of these samples, the frequency of systems with larger $q$ is much easier to measure and has a smaller required correction for biases.  There is still considerable disagreement as to the shape of the distribution function $f(q)$, but all of these distributions are consistent with\nocite{Latham2002}
\begin{equation}
\frac{1}{3}f(q=0.8)dq < f(q=0.1)dq < 3 f(q=0.8)dq.
\end{equation}
Duquennoy and Mayor (1991)\nocite{Duquennoy1991} find that 57\% of systems with F or G primaries are binary, while for comparison, Fischer and Marcy (1992)\nocite{Fischer1992} find that 42\% of systems with M dwarf primaries are binaries.  Taking $f(q)$ flat to within a factor of three as above, this means that 
\begin{equation}
\frac{N(\textrm{GM systems})}{N(\textrm{G stars})} \approx \frac{N(\textrm{FM systems})}{N(\textrm{F stars})} \sim 0.086^{+0.171}_{-0.057}.
\end{equation}
Additionally, from the Salpeter initial mass function \cite{Salpeter}, the number density of stars goes as $M^{-1.35}$.  Therefore, there should be approximately 4 times as many systems with M star primaries as G star primaries.  Thus, the fraction of M stars in binaries with an F or G mass companion is 
\begin{equation}
\frac{N(\textrm{GM systems})}{N(\textrm{M stars})} \sim 0.021^{+0.043}_{-0.014}.
\end{equation}
Finally, the maximum radius of an AGB star is around 1 AU.  So, our model requires a binary with a separation of order 1-2 AU.  The period distribution from Duquennoy and Mayor shows a near-Gaussian distribution in log(period) centered at $\textrm{log}(T)=4.8$ with $\sigma = 2.3$.  So, requiring a separation between 0.5 AU and 2 AU as an estimate of our requirements, we find that the final fraction of carbon M stars should be
\begin{equation}
\frac{N(\textrm{carbon M})}{N(\textrm{M})} \sim 0.0021^{+0.0043}_{-0.0014},
\label{eq:uncertainty}
\end{equation}
where this is again an underestimate of the uncertainty and only represents the uncertainty in the distribution function $f(q)$.

Finally, between 5\% and 15\% of AGB stars show C/O$>1$ \cite{Wallerstein1998}.  Current models of AGB stars indeed show that the carbon content in the stellar atmosphere increases throughout the lifetime of the AGB with additional thermal pulses, as does the mass loss rate.  If this is indicative of all AGB stars showing carbon but for only 5-15\% of their lifetime, then the above rate should be our final estimate of the frequency of carbon stars among the M dwarf population.  If not, then our estimate must be reduced accordingly (cf. Karakas \& Lattanzio 2007\nocite{Karakas2007}, Marigo 2002\nocite{Marigo2002}).  The SDSS measurement of the fraction of M dwarfs with C/O$>1$ is 0.3-0.4\% \cite{Schroeder2003}.  Therefore, if indeed the majority of AGB stars have a carbon phase, our model is consistent with the observed frequency.

\section{Additional Predictions}
\label{sec:GandS}

In the previous section, it was shown that the predicted frequency of carbon stars for our model is consistent with observations by SDSS.  However, there are still large uncertainties in not only the binary mass ratio distribution function $f(q)$, but also in the initial mass function.  An improvement in either of these would reduce the uncertainty in our final frequency of carbon M dwarfs to one much smaller than a factor of three, but at this point the predicted frequency is not by itself conclusive evidence for this model of their formation.  So let us consider other ways in which this model might be tested.

Perhaps the most striking prediction of a model in which carbon dwarfs are post-mass transfer binaries deals with the residual primary object.  This model predicts that most carbon dwarfs will have a white dwarf binary companion with a separation of order 1 AU.  If the AGB progenitor was an F or G star, the white dwarf will have a mass of order $0.5 M_\odot$, and therefore
\begin{equation}
\frac{v_{\textrm{{\small orb}}}}{\textrm{sin }i} \sim 15\textrm{ km/s}.
\end{equation}
Depending on the angle of inclination, it should be possible to resolve such systems using Doppler velocities.  Indeed, the radial velocity binary G77-61 \cite{Dearborn1986} and the spectroscopic binaries PG 0824+289 \cite{Heber1993} and CBS 311 \cite{Liebert1994} are known to be carbon dwarf-white dwarf systems.  To this point, the limitations on this search have been primarily sample size.  Downes et al. (2004)\nocite{Margon2004} find 107 carbon dwarfs with $15.6 < r < 20.8$.  It should be straightforward to observe these objects and search for a binary companion.  A larger sample of carbon dwarf-white dwarf systems would be compelling evidence for the binary mass transfer model.

Another test comes from considering the end state of a system containing an AGB progenitor and a secondary G star small enough that it has not yet left the main sequence.  Again using the Bondi-Hoyle approximation, the larger mass of the G star results in
\begin{equation}
\dot{M}_{dwarf} \approx -0.39\dot{M}_{AGB}.
\end{equation}
However, a G star with solar metallicity is not fully convective, and only the thin stellar envelope ($M\sim 0.02 M_\odot$) must be polluted.  Therefore, the final atmospheric abundances of carbon and oxygen will be very close to those of the carbon AGB donor.  Therefore, this model predicts a considerable carbon F and G star population in addition to the carbon M star population.

The fraction of carbon G stars is calculated using the techniques in {\S~\ref{sec:dynamics}}.  In this case, the initial mass function does not enter because we are looking at G-G binaries.  Further, the uncertainty in $f(q)$ in this range is not a factor of three, but rather closer to 50\% as \cite{Duquennoy1991}
\begin{equation}
\frac{N(\textrm{GG systems})}{N(\textrm{G primaries})} \approx 0.06 \pm 0.03.
\end{equation}
This results in
\begin{equation}
\frac{N(\textrm{carbon G})}{N(\textrm{G})} \approx 0.006 \pm 0.003.
\label{eq:predict}
\end{equation}
 Because $f(q)$ is known with much greater precision for systems with F/G secondaries, a good estimate of this population would provide a better test for the binary mass transfer model than the fraction of M stars with C/O$>1$.

Unfortunately, while it has been easy to distinguish carbon M stars in the SDSS color space because of the lack of TiO lines, there is no similarly distinctive feature of a carbon F/G star.  However, in the Downes et al. sample of carbon stars, a small fraction do show Balmer and Ca~{\small II} lines and are believed to be F or G carbon stars.  Approximately 20 more of these objects were found by searching for metal-poor F subdwarfs \cite{Schroeder2003}.  If this fractional abundance is consistent with the prediction given by Eq. \ref{eq:predict}, it would be very strong confirmation of the post-mass transfer model.

\section{Summary}
\label{sec:conclude}
We have used new observational evidence from the Sloan Digital Sky Survey to re-examine the model of de Kool \& Green (1995)\nocite{DeKool1995} that dwarf carbon stars are formed via mass transfer from a carbon AGB companion.  Using the SDSS composite spectrum of these stars, we have shown that carbon stars of M spectral type indeed are stars with C/O$>1$ and that their masses are consistent with main sequence M stars.  Despite large uncertainties (see \S~\ref{sec:dynamics}), we show that the predicted fraction of M stars with C/O$>1$ is consistent with the observed fraction.  This agreement is a requirement of our model but not by itself sufficient proof of its veracity.

Two additional predictions should prove testable in the near future.  There are already hints that many carbon dwarfs may be in binaries with white dwarfs.  Now that there is a sample of over 100 carbon M stars, it should be possible to search for Doppler velocities we predict may be as large as $15\textrm{ km/s}$.  Such a search could likely be completed within two years.  The other would be a search for F/G carbon stars.  Large uncertainties in the mass ratio distribution of binaries and the initial mass function apply much more strongly to M stars than F/G stars.  As a result, agreement between the theoretical and observed abundance of F/G carbon stars would be strong evidence for our model.  While F/G carbon stars do not lie in quite as distinctive a region of the color space produced by the Sloan Digital Sky Survey, there is already a project underway to search for F/G carbon stars, and this project should be able to provide a fairly stringent test for this model of carbon star formation upon completion.  So, it should be possible for this mass-transfer model to be either verified or disproven in the near future, and without the time-consuming new observations required by a search for white dwarf companions.


If this model for carbon dwarf formation is confirmed, we should turn our attention to its implications.  One intriguing set of objects are the two very low metallicity carbon dwarfs discussed in Cohen et al. (2004)\nocite{Cohen2004}.  According to this model, a metal-poor carbon M dwarf is the endstate of a binary system with a metal-poor AGB and a metal-poor M dwarf.  A comparison of the abundances of the carbon M dwarf to those of an M dwarf of similar metallicity would allow a calculation of the elemental abundances present when its companion was a carbon AGB.  Such a calculation has already been done with a metal-poor turnoff star to find abundances of its former AGB companion \cite{Aoki2002,Sivarani2003}, so it should be feasible using a carbon M dwarf as well.  Therefore, one of the implications of this model is that metal-poor carbon M dwarfs offer a probe into the elemental abundances of the earliest massive stars while they were on the asymptotic giant branch.  While the main sequence abundances of these stars are of greater interest, particularly as a probe of primordial nucleosynthesis, these cannot be obtained without a considerable improvement in our understanding of the changes in elemental abundance between the main sequence and the asymptotic giant branch. 

\acknowledgements
The authors would like to thank Paul Green, Gillian Knapp, and Bob Kurucz for valuable consultations during this project.  CLS is supported under a National Science Foundation Graduate Research Fellowship.

\end{document}